\input harvmac
\input epsf

\newcount\figno
\figno=0
\def\fig#1#2#3{
\par\begingroup\parindent=0pt\leftskip=1cm\rightskip=1cm\parindent=0pt
\baselineskip=12pt
\global\advance\figno by 1
\midinsert
\epsfxsize=#3
\centerline{\epsfbox{#2}}
\vskip 14pt

{\bf Fig. \the\figno:} #1\par
\endinsert\endgroup\par
}
\def\figlabel#1{\xdef#1{\the\figno}}
\def\encadremath#1{\vbox{\hrule\hbox{\vrule\kern8pt\vbox{\kern8pt
\hbox{$\displaystyle #1$}\kern8pt}
\kern8pt\vrule}\hrule}}

\overfullrule=0pt

\noblackbox
\parskip=1.5mm

\def\Title#1#2{\rightline{#1}\ifx\answ\bigans\nopagenumbers\pageno0
\else\pageno1\vskip.5in\fi \centerline{\titlefont #2}\vskip .3in}

\font\caps=cmcsc10

\noblackbox
\parskip=1.5mm

  
\def\npb#1#2#3{{\it Nucl. Phys.} {\bf B#1} (#2) #3 }
\def\plb#1#2#3{{\it Phys. Lett.} {\bf B#1} (#2) #3 }
\def\prd#1#2#3{{\it Phys. Rev. } {\bf D#1} (#2) #3 }
\def\prl#1#2#3{{\it Phys. Rev. Lett.} {\bf #1} (#2) #3 }
\def\mpla#1#2#3{{\it Mod. Phys. Lett.} {\bf A#1} (#2) #3 }

\def\cmp#1#2#3{{\it Commun. Math. Phys.} {\bf #1} (#2) #3 }

\def\bb#1{{\tt hep-th/#1}}

\def\jhep#1#2#3{{\it J. High Energy Phys.} {\bf #1} (#2) #3 }


           \def\CO{{\cal O}} 
   
\def\CL{{\cal L}}   
   
\def\CM{{\cal M}}  
\def\CN{{\cal N}} \def\CS{{\cal S}}


\def\dj{\hbox{d\kern-0.347em \vrule width 0.3em height 1.252ex depth
-1.21ex \kern 0.051em}}

\def\Tr{{\rm Tr\,}}

\def\ket{\rangle}
\def\bra{\langle}

\def\ox{\bar x}

\def\Dirac{\,\raise.15ex\hbox{/}\mkern-13.5mu D}
\def\dirac{\,\raise.15ex\hbox{/}\kern-.57em \partial}
\def\shalf{{\ifinner {\textstyle {1 \over 2}}\else {1 \over 2} \fi}} 
\def\sshalf{{\ifinner {\scriptstyle {1 \over 2}}\else {1 \over 2} \fi}} 
\def\sfourth{{\ifinner {\textstyle {1 \over 4}}\else {1 \over 4} \fi}}

\lref\rlargen{G. 't Hooft, \npb{75}{1974}{461.}}

\lref\rmaldacobi{N. Itzhaki, J. Maldacena, J. Sonnenschein and S. Yankielowicz,
\prd{58}{1998}{046004} \bb{9802042.}
}

\lref\rads{J. Maldacena, {\it Adv. Theor. Math. Phys.} {\bf 2} (1998)
231 \bb{9711200.}  
 S.S. Gubser, I.R. Klebanov and A.M. Polyakov, 
\plb{428}{1998}{105} \bb{9802109.} E. Witten,
{\it Adv. Theor. Math. Phys.} {\bf 2} (1998)
253 \bb{9802150.} }

\lref\rHP{G.T. Horowitz and J. Polchinski, \prd{55}{1997}{6189}
\bb{9612146.}}

\lref\rrecipe{M. Berkooz, A. Sever and A. Shomer, \bb{0112164.} 
}
\lref\rrewit{E. Witten, \bb{0112258.}} 

\lref\rotros{P. Minces and V.O. Rivelles, \jhep{0112}{2001}{010} \bb{0110189.}
W. Muck, 
\plb{531}{2002}{301}  
\bb{0201100.} 
P. Minces,  
\bb{0201172.} 
A.C. Petkou,  
\bb{0201258.}    
E.T. Akhmedov,  
\bb{0202055.} 
A. Sever and  A. Shomer,  
\bb{0203168.}} 

\lref\rwitmast{E. Witten, {\it ``Recent Developments in Gauge Theories"}, 1979
Carg\`ese Lectures. Ed. G. 't Hooft et. al.,  Plenum, New York,  (1980).} 

\lref\rexodef{O. Aharony, M. Berkooz and E. Silverstein, \jhep{0108}{2001}{006}
\bb{0105309.} \bb{0112178.}} 

\lref\rozhas{A. Hashimoto and Y. Oz, \npb{548}{1999}{167}  
\bb{9809106.}}  
 
\lref\rthresholds{J.L.F. Barb\'on, I.I. Kogan and E. Rabinovici,
\npb{544}{1999}{104} \bb{9809033.}}

\lref\rwitbar{E. Witten, \npb{160}{1979}{57.}}

\lref\rHPage{S.W. Hawking and D. Page, \cmp{87}{1983}{577.}}

\lref\rwitheta{E. Witten, \prl{81}{1998}{2862}  
\bb{9807109.}}  

\lref\rwitthp{E. Witten, {\it Adv. Theor. Math. Phys.} {\bf 2} (1998)
505 \bb{9803131.}}

\lref\roldmm{S.R. Das, A. Dhar, A.M. Sengupta and S.R. Wadia, \mpla{5}{1990}{1041.}
L. Alvarez-Gaum\'e, J.L.F. Barb\'on and C. Crnkovic, \npb{394}{1993}{383.} G. 
Korchemsky, \mpla{7}{1992}{3081,} \plb{256}{1992}{323.} I.R. Klebanov, 
\prd{51}{1995}{1836.} I.R. Klebanov and A. Hashimoto, \npb{434}{1995}{264.}
J.L.F. Barb\'on, K. Demeterfi, I.R. Klebanov and C. Schmidhuber, \npb{440}{1995}{189.}}
 

\baselineskip=15pt

\line{\hfill CERN-TH/2002-112}
\line{\hfill {\tt hep-th/0206207}}

\vskip 1.0cm

\Title{\vbox{\baselineskip 12pt\hbox{}
 }}
{\vbox {\centerline{Multitrace AdS/CFT and Master Field Dynamics  }
\vskip10pt
\centerline{}
}}

\vskip0.5cm

\centerline{$\quad$ {\caps 
J.L.F. Barb\'on\foot{ On leave
from Departamento de F\'{\i}sica de Part\'{\i}culas da 
Universidade de Santiago de Compostela, Spain.}  
}}
\vskip0.5cm

\centerline{{\sl  Theory Division, CERN, 
 CH-1211 Geneva 23, Switzerland}}
\centerline{{\tt 
barbon@cern.ch}}

\vskip0.5cm

\centerline{\bf ABSTRACT}

 \vskip 0.2cm

 \noindent 

We consider   gauge theories with 
multitrace deformations
in the context of certain  AdS/CFT
models with explicit breaking of conformal symmetry and supersymmetry.
In particular, we study the standard four-dimensional confining model based on
the D4-brane metric at finite temperature. 
We work in the self-consistent Hartree approximation, which becomes exact
in the large-$N$ limit and is equivalent to the  AdS/CFT 
multitrace prescription that has been proposed in the literature.
We show that generic 
multitrace perturbations
have important effects on the phase structure of these  models. Most
notably they can induce  new types of large-$N$  
first-order phase transitions.

\vskip 0.5cm

\Date{June 2002}
               

\vfill





\baselineskip=15pt

\newsec{Introduction}

\noindent

In 't Hooft's large-$N$ limit of gauge theories \refs\rlargen, the
 scaling of the
bare gauge coupling $g^2 \sim 1/N$ is 
tuned so that the vacuum energy is proportional to $N^2$.  
This  scaling generalizes to an arbitrary action  according to the rule:   
\eqn\ac{
S= N^2 \,W(\CO_1, \CO_2, \dots)\;,}
where $W$ is a general functional of operators of
the symbolic form  
\eqn\defoi{
\CO_n = {1\over N}\,\Tr \,F^n \;,}
the set of
 single-trace 
gauge-invariant operators with expectation values of $O(1)$ in the
large-$N$ limit\foot{We shall not discuss here operators with anomalous
large-$N$ scaling. The most notable example is the theta-term with 
scaling  
$\bra \,\Tr\,F\wedge F\,\ket = O(1)$.}.
 For more general theories, including scalar fields and
fermions in the adjoint representation, we extend the basic family of
gauge-invariant operators to include these fields as well. 
These operators become quasi-classical in the large-$N$ limit, in the
sense that 
$$
\lim_{N\to \infty} \;\bra \,\CO \,\CO'\,\ket =\lim_{N\to\infty}\;
 \bra\,\CO\,\ket \;\bra\,
\CO' \,\ket\;.
$$
 This means that there is a notion of saddle-point
configuration $-$a ``master field" defined up to gauge transformations,
which makes the $1/N$ expansion into a semiclassical expansion
 \refs\rwitmast. 

Known or conjectured master fields are usually established for 
single-trace actions, i.e. for linear $W$ in \ac, 
 such as the  Yang--Mills action.
 However,  
the behaviour of master fields under perturbations by multitrace operators
is of primary interest, 
especially in the context of the AdS/CFT correspondence \refs{\rads}.
 In the holographic mapping, multitrace
 operators are associated to multiparticle states in the bulk theory. Hence
they correspond to exotic deformations of the string background \refs\rexodef.
Moreover, truly non-perturbative effects  in the bulk theory 
 manifest themselves
as finite-$N$ multitrace  effects on the CFT. 
 This
is simply the translation of the fact that only $O(N)$ elementary powers
of the form $\Tr F^n$ are algebraically independent: for $n\gg N$ the
single-trace operator decomposes as a sum of products of lower-order
single-trace operators. Hence, the  
spectrum of the  bulk theory must deviate significantly from a Fock space
for states with $O(N)$ ``particles".   

It is then very interesting to study the effect of multitrace deformations
on the AdS/CFT saddle point, particularly the effect of deformations that
are   non-polynomial
 in the traces.   
Recently, the  AdS/CFT algorithm was modified to incorporate multitrace
operators \refs{\rrecipe,\rrewit}\ (see also \refs\rotros).
 Here we elaborate on some points made in \refs\rrewit\ to
 argue that this modification can be understood in
rather general terms, as an application of the mean-field approximation.

In analysing  the large-$N$  master field,  
we could  attempt a saddle-point approximation once we have managed to exactly
 integrate out
 $O(N^2)$ degrees of freedom. If we remain with $O(N)$ degrees of freedom,
this sets the order of magnitude of the fluctuations. Since the action is
of $O(N^2)$, we have a sharp saddle point. In practice, such a program only
works in very restricted models in low dimensions, where we can integrate
out explicitly the $O(N^2)$ angular variables (for a discussion of multitrace
operators in these models, see \refs\roldmm). 
Still, one can argue in great generality that in the leading large-$N$
 approximation  $W$ can be taken essentially linear.  
 
Let us suppose that we have managed to change variables in the path integral
from the gauge field $A_\mu$ to the set of gauge-invariant monomials $\CO_n$
with $n< O(N)$. In the process we generate a complicated 
(non-local) effective action $\Gamma$. 
At the large-$N$ saddle-point we have: 
\eqn\saddle{
{\partial \Gamma \over \partial \CO_n} ( \CO_{cl}) + 
{\partial W \over \partial \CO_n} ( \CO_{cl}) =0 
\;,} 
where we have incorporated the fact that the solution of the saddle-point
equations is nothing but the planar expectation values: $\CO_{cl} \equiv
\lim_{N\to \infty} 
\bra \CO_n \ket$. 

 In view of \saddle, it is clear that these equations are exactly the same
as those that follow from a model with a single-trace action given by
\eqn\str
{{\overline W}(\CO) = \sum_n \;{\overline \zeta}_n \;\CO_n\;,}
where the effective single-trace couplings ${\overline \zeta}_n$ are
given by
\eqn\mastereq{
{\overline \zeta}_n =  
{\partial W \over \partial \CO_n} ( \CO_{cl})  
\;.}
Therefore, provided we only consider the planar $N\rightarrow \infty$ limit,
any quantity of the original theory \ac\ can be computed in the single-trace
theory \str, with the expectation values $\bra \CO \ket$ being determined
self-consistently.\foot{This argument assumes some explicit regularization,
so that the path integral measure is defined  
 independently of the details of the action.}
 Thus, in the AdS/CFT set-up, the combination
$
\partial W (\CO_{cl})/ \partial \CO_n 
$ 
plays the role of the source for the single-trace operator $\CO_n$, and this
precisely determines  the  boundary conditions proposed in \refs{\rrecipe,
\rrewit}.  

Our discussion shows that the basic phenomenon is more general than the
particular AdS/CFT set-up. Namely, it is a general consequence of the fact
that the Hartree (or Thomas--Fermi) approximation becomes exact in the
large-$N$ limit (see for example \refs\rwitbar).
 In this limit, the interactions between the gauge-invariant
variables $\CO_n$ can be substituted by the interaction of each variable
with a collective mean field that must be determined sefl-consistently. 

We should emphasize that these rules are only valid in the strict $N=\infty$
limit. The $1/N$ corrections will alter the master equation \mastereq\ since the
Hartree approximation obtains corrections. Equivalently,  
the AdS/CFT boundary conditions of \refs{\rrecipe,\rrewit}\ will receive $1/N$
corrections, in addition to the usual loop corrections in the bulk of AdS. 

\newsec{Master Field Dynamics} 

\noindent

To be more specific, let us suppose that the deformations by 
 a certain  single-trace operator $\CO$: 
\eqn\addd{
\delta \,S =N^2 \;\zeta \, \int d^d x\; \CO
\;,}
are under control, in the sense that we  are able to compute 
the planar one-point function $\bra \CO\ket_\zeta$ as 
a function of $\zeta$ and the other couplings of the Lagrangian. Then 
we can compute any planar expectation value of the  
more general theory with perturbation
\eqn\newp{
N^2 \,\int d^d x \,\mu^d \,F(\mu^{-d_\CO}\,\CO)\;,}
where $F$ is general function, $\mu$ is a mass scale and $d_\CO$ is the
scaling dimension of the operator $\CO$ in the single-trace model.
 We simply do our calculations   in the single-trace theory
$\addd$ with perturbation
\eqn\stt{
\delta\,{\overline S} =N^2 \;{\overline \zeta} \,\int d^d x\; \CO\;,}
where ${\overline \zeta}$ is given self-consistently  by the  solution
of the ``master equation":  
\eqn\sefl{
G(\,{\overline \zeta}\,) \equiv {\overline \zeta} - \mu^{d-d_\CO} \;
F'  \left[\,\mu^{-d_\CO}\,\bra \CO\ket_{\overline \zeta}\,\right] =0\;,} 
where the prime denotes differentiation.
 In principle, we can give ${\overline \zeta}$
 a space-time dependence so that \sefl\ becomes a functional equation for an
effective source. Such a generalization is appropriate to compute correlation 
functions in the multitrace-deformed theory.  However, for the purposes
of this paper we are only interested in vacuum properties of the master
field, i.e. we consider only condensates and effective couplings that
are translationally invariant in ${\bf R}^d$.  

The ``master equation"  \sefl\ implies
 that multitrace deformations whose single-trace
``elementary" operator has a 
vanishing one-point function are equivalent (in the large-$N$
limit)  to
single-trace deformations, i.e. a constant shift
$$
\delta\,\zeta = \mu^{d-d_\CO} \;F' (0)
$$
of the coupling dual to the single-trace operator $\CO$.   
 Hence, in order to have specifically  
 new phenomena associated to multitrace deformations we need
non-vanishing  condensates. This means that   
 the auxiliary single-trace model with perturbation
\stt\ must break conformal invariance either explicitly or spontaneously.
 Since the one-point function depends on the particular state
that we are considering, it is plain
 that the physical properties of multiple-trace
deformations have a strong dependence on the full physics of condensates of
the associated single-trace model.

We may take $F$ as a non-polynomial function of single-trace
operators. However, we implicitly treat the non-linear terms as a perturbation
since the scaling dimensions $d_\CO$ are defined with respect to the
single-trace theory. At any rate, it is interesting to evaluate \sefl\ 
when the function $F$ becomes non-polynomial.

Our main observation in this paper is the following. 
The function $G(\,{\overline \zeta}\,)$ may have a complicated structure, being
non-linear in both ${\overline \zeta}$ and the couplings of the
bare Lagrangian $W$. 
In particular, if $G(\,{\overline\zeta}\,)$ has various nodes, we have
a set of solutions $\{\,{\overline \zeta}_\alpha\,\}$
 for a given fixed value of
the microscopic couplings in $W$. In this case we must select the
master field that dominates the large-$N$ dynamics among the various
solutions ${\overline \zeta}_\alpha$.  

By analogy with similar situations in large-$N$ physics  we characterize the
dominating master field by requiring that the partition function  be 
maximized:
\eqn\maxze{
\lim_{N\to \infty}\; {1\over N^2} \,\log\;Z_W = {\rm max}_{\;\alpha} \;\left[
\;\lim_{
N\to\infty} \;{1\over N^2}\, \log\;
 Z(\;{\overline\zeta}_\alpha\;)_{\overline W}
\;\right]\;.}
Large-$N$ phase transitions induced by the multitrace couplings will
arise when the dominating zero of $G(\,{\overline \zeta}\,)$ changes
discontinuously as
a function of the microscopic couplings in $W$. These phase transitions
will be characterized by a ``latent heat" release of $O(N^2)$.  
Typically, $Z(\,{\overline\zeta}\,)_{\overline W}$ will be  a monotonic 
function of ${\overline\zeta}$, so that a change of branch in \maxze\
will require that the cardinality of the solution set $\{\,{\overline
\zeta}_\alpha\,\}$ changes as a function of $W$. 

Although the phenomena described so far are expected to be rather general,
we will illustrate them in a specific example in the context of the AdS/CFT
correspondence.

\newsec{Multitraces in Deformed QCD}

\noindent

  As a concrete example along the previous lines,
we  consider a  regularized version of four-dimensional 
non-supersymmetric Yang--Mills theory  that has been introduced in \rwitthp. 
 In its most straightforward definition,
the model  is given by the low-energy theory on the world-volume
of a stack of $N$ parallel 
D4-branes at finite temperature $T$. Equivalently, we can view it
as a Scherk--Schwarz compactification of the D4-branes on ${\bf S}^1 \times
{\bf R}^4$,   the  compact
circle having  size $1/T$.  At large distances on ${\bf R}^4$ the effective
theory is a four-dimensional Yang--Mills theory modified at energies of
$O(T)$ by remnants of
the five-dimensional $\CN=4$ super Yang--Mills theory. The action is given by  
\eqn\und{
{N \over g_{\rm YM}^2} \int \,d^4 x\; \CL = {1 \over 4g_{\rm YM}^2} \int
d^4 x\; \left(\Tr F^2 + \dots\right)\;,}
where the dots stand for other fields such as fermions and scalars of the
parent $\CN=4$ theory, suppressed by powers of the cutoff scale $\mu =T$.   
Planar quantities are  functions of the 't Hooft coupling  $\lambda$,  
 defined at the cutoff scale $\mu$. In terms of the microscopic
parameters of the parent D4-brane theory  we have
$$
\lambda \equiv  g_{\rm YM}^2 (\mu)\, N  \sim g_s\,N\,\mu\,\sqrt{\alpha'}\,,
$$
where $g_s$ is the string coupling and $\alpha'$ is the string's Regge slope. 
For  $\lambda  \ll 1$ we have the standard planar perturbation theory
of the four-dimensional Yang--Mills theory. On the other hand, for $\lambda
\gg 1$ we have a good description in terms of the low-curvature expansion of the black
D4-brane metric. In this case, the expansion parameter  
is controlled by the curvature of the near-horizon metric in string
units:
$$
{\alpha' \over R_{c}^2} \sim {1\over \lambda} \,, 
$$
with $R_c$ the curvature radius. Defining 
$$
x\equiv {1\over \lambda},
$$
the supergravity description is good for $0<x\ll 1$. At $x\sim 1$ we have
the standard ``correspondence point" in the sense of \refs\rHP, which represents
the matching to the perturbative regime. As long as we only look at energy
scales of $O(1)$ in the large-$N$ limit, we can neglect non-perturbative
thresholds associated to large values of the dilaton, since these involve
explicit powers of $N$.

The simplest multitrace perturbation in these models is 
 a non-linear function  of the Lagrangian density, 
\eqn\plag{
S= {N^2 \over \lambda} \int \CL + N^2  \,\int \,\mu^{4}\; F(\mu^{-4}\;\CL)\;.}
 According to \sefl, all physical quantities  in this model, 
such as thermodynamic functions, condensates,
Wilson loops, etc. ,  can be computed in the large-$N$ limit in the auxiliary  
undeformed model
\eqn\und{
{\overline S} = {N^2 \over {\overline \lambda}} \int \CL \;,}
with effective 't Hooft coupling ${\overline \lambda}$  
 given by the solution of the equation
\eqn\sol{
{\overline \lambda}^{\,-1} = \lambda^{-1} + F' \left[\,\mu^{-4}
\,\bra  \,
\CL \, \ket_{\overline \lambda}\,\right]\;,}
where the gluon condensate $\bra \CL \ket_{\overline \lambda}$ is determined
by the expectation value of the action:  
\eqn\opf{
\left\bra \;\int\,\CL\;\right\ket_{\overline \lambda}={\rm Vol}\,(\,{\bf R}^4\,)
\;\left\bra 
 {1\over 4N}\,\Tr \,F^2 +
 \dots \right\ket_{{\overline \lambda}} = {1\over N^2} \;
{\overline \lambda}^{\;2} \,{\partial \over \partial {\overline \lambda}}
 \,\log\,Z(\,{\overline \lambda}\,)\;.}
The partition
 function in the planar supergravity approximation
is defined in terms of the thermal free energy of the D4-brane (see,
for example \refs\rthresholds): 
\eqn\pps{
{1\over  {\rm Vol}({\bf R}^4)}\;\log\;Z(\,{\overline \lambda}\,) = 
N^2 \,C\;{\overline \lambda}
 \,\mu^{4}\; ,}
where   $C$ is a 
 positive numerical constant. This expression for the partition function
has been normalized to the Euclidean action of the wrapped D4-brane metric
with supersymmetric boundary conditions, i.e. we define the five-dimensional
thermal free energies  with respect to the $T=0$ vacuum. 

 Notice that, even if
the general multitrace deformation of the $T=0$ D4-brane theory may break supersymmetry,
the $N=\infty$ effective theory \und\ {\it does not}. Hence, the D4-brane theory reduced
on a supersymmetric circle will be supersymmetric at $N=\infty$ and no condensates
will be induced.\foot{Casimir energies are not induced either,  since the
D4 world-volume is flat  ${\bf S}^1 \times {\bf R}^4$.}
  This implies that the condensates are  
entirely due to thermal effects of the D4-brane theory and our normalization of
\pps\ is the physically correct one.

Combining \opf\ and \pps\ we find the value of the gluon condensate 
(c. f. \refs\rozhas): 
\eqn\tras{
\mu^{-4}\;\left\bra 
 \,\CL \,\right\ket_{\overline \lambda}
 = C\;{\overline \lambda}^{\;2 }
 \;.}
This expectation value has the crucial property of diverging as  
${\overline\lambda} \rightarrow \infty$. 
Since this is precisely the supergravity regime of the effective
single-trace theory, we learn that multitrace deformations are potentially
stronger in the region where AdS/CFT is under quantitative control and
they may be reliably studied.

In terms of  the dimensionless expansion parameters  $x\equiv 1/\lambda$ and
$\ox \equiv 1/{\overline \lambda}$ the master equation reads  
\eqn\edim{
G(\ox) \equiv \ox - x - F' \left[ C/\ox^{\;2}\right] =0.}  
Equation \edim\ was derived within the supergravity approximation to
the  near-horizon black D$4$-brane solution. In terms of the supergravity  
expansion  parameter $\ox$, this is the regime:
\eqn\les{
0< \ox \ll 1\;.}
 As before, 
 these limits ignore other thresholds that are
 related to large dilaton corrections and are of subleading order in the $1/N$
expansion.

One important property of \edim\ is the redundancy of the description in
terms of the original variables in the microscopic Lagrangian, i.e. the
coupling $x$ and the multitrace couplings that define the function $F$.
 For a fixed value
of $\ox$ all models in the codimension-1 submanifold  
$$
\CM_{\ox} : \;\;\;\;x-\ox + F' =0
$$
 have the same
large-$N$  properties. 
The region of the microscopic coupling space  
 where supergravity is a good approximation
 is the union of these submanifolds for $0<\ox <1$:  
\eqn\sureg{
\CS = \bigcup_{0<\ox <1} \;\;\CM_{\ox} \;.}
One component of the boundary is  $\CM_{\ox =0}$ defined as
\eqn\lines{
\CM_0: \;\;\;\;x+ F'(\infty)=0
\;.}
It
yields the strong-coupling (low-curvature) limit of the AdS/CFT background.
On the other hand, the correspondence line (the
matching to perturbative variables)
occurs at $\ox =1$ or
\eqn\linec{
\CM_1:\;\;\;\;x+F'(C) -1=0\;.
}
Although  $\CM_0 \cup \CM_1$ are components of
the boundary of $\CS$, they do not exhaust it in general.

\subsec{Multicritical Behaviour} 

\noindent

For $0<x\ll 1$ there is always a standard solution of  
\edim\ that is valid for very small multitrace couplings. This solution
has $\ox \approx x$ and can be obtained iteratively as the limit
of the set $\{\ox_{(k)}\}$ with
\eqn\iter{
\ox_{(k+1)} = x+ F'\left[ C/(\ox_{(k)})^2 \right]\;, \qquad \ox_{(0)} = x\,.}
However, it is clear that there will be other solutions if $F'[C/\ox^2]$
shows ``bumps"  in the supergravity interval $0<\ox < 1$.   

Let us assume that $F'$ admits a 
finite
 Laurent expansion around the origin, so that  the master equation takes
the form:
\eqn\mastere{
G(\ox;\,x,\, f_j) = \ox - x - \sum_{j\neq 0} {f_j \over \ox^{\,2j}} = 0\;.}
The $j=0$ term is equivalent to a constant shift of $x$ and has been removed
from \mastere. 

Our first result is  a simple consequence of the divergence 
of \tras. The pole part of $F$, corresponding
to $j<0$ in \mastere, has no dramatic effects  in the supergravity
interval $0<\ox \ll 1$.  Thus, 
 multitrace deformations that are completely
singular in perturbation theory become rather tame in the supergravity
approximation. This looks surprising at first sight, but it fits naturally
with the character of AdS/CFT as a strong/weak coupling duality with
respect to the 't Hooft coupling.   

Conversely, perturbations that are polynomial in multitraces translate into
non-analytic contributions to $G(\ox)$ and therefore dominate the supergravity
regime at  $\ox \rightarrow 0$. 
In this limit $G(\ox)$ diverges with a sign that is correlated with that
of $f_J$, $J$ being the largest value of the index $j$. In particular, for
$f_J <0$ and small there is always a solution:
\eqn\smalls{
\ox_- \approx \left(-{f_J \over x} \right)^{1\over 2J}\;.}
This solution disappears for $f_J >0$, unless one also dials the microscopic
't Hooft coupling to negative values: $x<0$.

We have found that for $0<x<1$ and small $|f_J|$, we have a discrete jump
in the number of solutions of the master equation as $f_J$ crosses  
 zero. This is a source of possible phase transitions.
 
A more general multicritical behaviour in the vicinity of $\ox \approx 0$
will depend on the higher multitrace powers.   Let us consider a
simple example of a deformation proportional to
\eqn\exdef{
{g_1 \over 2} \int \mu^{-4} \;\left(\Tr F^2 \right)^2  + {g_2 \over 3N} \int
\mu^{-8} \;\left(\Tr F^2 \right)^3\;.
}
 The master equation \mastere\ reads:
$$
G(\ox;\, x,\, f_1,\, f_2) = \ox - x- {f_1 \over \ox^2} - {f_2 \over \ox^4}=0\;,  
$$
with $f_i \sim g_i$ up to numerical constants.
Besides the standard solution $\ox_+ \approx x$ for very small $f_j$ there
are other interesting solutions. Consider $x>0$,  $f_2 >0$ and $f_1 <0$, with
$f_2  \ll |f_1 | \ll x$ and furthermore $x\,f_2 \ll f_1^2$. Then, the master    
equation has  two  small solutions in the vicinity of  
$$
\ox_- \sim \sqrt{-f_2 / f_1}, \qquad \ox'_- \sim  \sqrt{-f_1/x}
\;.$$ 
These solutions coincide for $f_1^2  \sim x\,f_2$ and disappear for
larger values of $f_2$. 

\subsec{Phase Transitions}

\noindent

The previous discontinuities in the solution set  of the master equation
translate into large-$N$ phase transitions. Since the partition function
scales as
\eqn\scalp{
\log\;Z(\,\ox\,) \propto {N^2 \over \ox}\;,
}
 we find that the dominant solutions in the supergravity approximation
 are those with the {\it smallest} value  of $\ox$ within the unit interval.    
The jump of the effective action across the transition from $\ox_\alpha$
to $\ox_\beta$ is given by
\eqn\jumb{
{1\over {\rm Vol} ({\bf R}^4)}\;
\log\;\left[{Z(\,\ox_\alpha\,) \over Z(\,\ox_\beta\,)}\right] = 
N^2 \,C\,\mu^4 \,\left({1\over \ox_\alpha} - {1\over \ox_\beta}\right)\;.}

Coming back to the examples in the previous subsection,
 we see that there is always a phase transition when $f_J$ crosses
zero from negative to positive
 values. In this case $\ox_\alpha =0$ and $\ox_\beta \approx x>0$. The
density of  ``latent heat"  in \jumb\ is infinite. This phase transition
is not hard to interpret. Since $f_J$ is the coupling of the multitrace
interaction of highest order, it dominates the limit of large field-strengths.
Hence, the very strong singularity for $f_J \rightarrow 0^-$ reflects
the fact that the microscopic action is not bounded below for $f_J <0$.  

A more physical phase transition with finite ``latent heat" takes place
in the two-coupling model 
 \exdef\ with $x>0$,  $f_2 >0$ and $f_1 <0$,
 when the two solutions around $\ox_- \sim \sqrt{-f_1 /x}$  
coalesce as we decrease the magnitude of $|f_1|/f_2$.  For small
values of this ratio the only solution is $\ox \approx x$. 

This example illustrates the general pattern of phase transitions in
this class of models.  When the minimal solution $\ox_-$ of the master equation
is separated from the first subleading one $\ox'_-$ by a local
maximum of $G(\ox)$, a variation of the parameters can bring the maximum
to zero and make the two solutions coalesce $\ox_- = \ox'_-$. A 
further variation of the parameters can  bring  the maximum to negative values
 and
make the double solution disappear. This generic situation is depicted
in Fig. 1 below.   

\fig{\sl 
A depiction of a typical phase transition. The solid line shows the 
function $G(\ox)$ with three zeros, $\ox_- < \ox'_- < \ox_+$. 
The dotted line shows the degeneration of the lower zeros $\ox_- = \ox'_-$
and their disappearance in favour of $\ox_+$. When the partition function
is dominated by the smallest solution this degeneration yields a 
large-$N$ phase transition.}{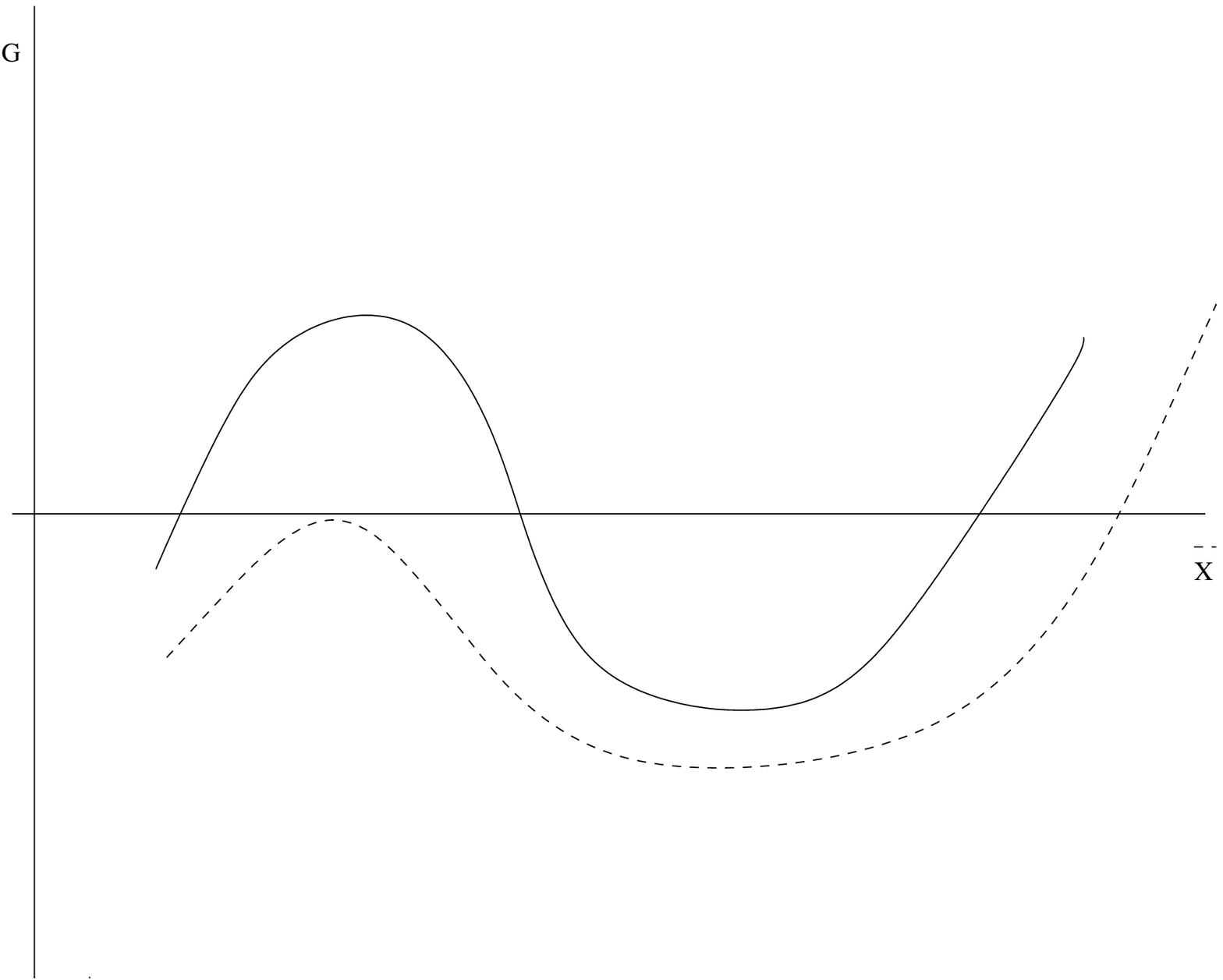}{5truein}

\newsec{Generalization to Other Dimensions} 

\noindent

This set-up can be  generalized to  the regularized Yang--Mills model
on ${\bf R}^p$, with $p<5$,  in terms of a hot D$p$-brane model and the
 corresponding
generalization of the  AdS/CFT correspondence
  \refs\rmaldacobi. In this case, the effective dimensionless  't Hooft coupling
normalized at the cutoff scale $\mu=T$ is given by
$$
\lambda_p (\mu) \sim  g_s\,N\,(\,\alpha'\,)^{\,p-3 \over 2} \;\mu^{\,p-3}\;.
$$
This is the expansion parameter of the planar perturbative expansion.
The expansion parameter of the supergravity approximation that arises at
$\lambda_p (
\mu) \gg 1$ is:
\eqn\oext{
x\equiv \left({1\over \lambda_p (\mu)} \right)^{1\over 5-p} \sim {\alpha' \over R_{
c}^2}\;.
}
The large-$N$ solution of these models perturbed by  multitrace interactions
of the form
\eqn\pertp{
N^2 \;\int d^p x \;\mu^p \;F\left(\,
 \mu^{-4} \; \CL  \;\right)
}
can be studied along lines  similar to the $p=4$ case above. Here, $\CL$ denotes
the Yang--Mills Lagrangian operator, corrected by regularization artefacts
at the scale $\mu=T$.  As before, one reduces the
problem to the study of an effective single-trace model with supergravity
expansion parameter
\eqn\defex{
\ox \equiv \left(1/ \,{\overline \lambda}_p (\mu)\right)^{1\over 5-p}} 
that is determined self-consistently. The supergravity regime of the
$N=\infty$ problem is then given by $0<\ox \ll 1$. 
 The partition function in the single-trace model with effective coupling
$\ox$ is 
\eqn\partp{
{1\over {\rm Vol} ({\bf R}^p)} \;\log \;Z(\,\ox\,) = N^2 \; (5-p)\,C_p\;
\mu^p \;\ox^{\,3-p}\;,}
where $C_p$ is a positive numerical constant. The gluon  condensate
is given by 
\eqn\lagp{
\mu^{-4} \;\bra \,\CL\,\ket_{\ox} = (p-3)\,{C_p\over \ox^{2}}\;.}
These expressions show that the $p=3$ case, based on the hot D3-brane,
yields trivial multitrace deformations in this approximation. This is
a consequence of the free energy of D3-branes being very smooth for
large 't Hooft coupling.    
 Of course, this situation changes when considering
subleading terms in the $\alpha'$ expansion of the supergravity background.
It is interesting to study these corrections in more detail, although
we will not attempt to do this here.

For $p<3$ one finds a situation somewhat similar to that discussed before
in the $p=4$ case. The master equation for $\ox$ reads:
\eqn\masterp{
\ox^{\,5-p} = x^{\,5-p} + F'\left[\,(p-3)C_p /\ox^{\,2}\,\right]=0\;.}
Hence, the same qualitative properties follow, regarding the multiplicity
of solutions  
at small $\ox$. In particular, the crucial
singularity at $\ox =0$ of the gluon condensate \lagp\
  still holds.  

The main difference with $p=4$ is that, according to \partp, for $p<3$ it is
the {\it largest} solution $\ox_+$ that dominates the partition function.
As a result, we expect that the standard
solution $\ox \approx x$ will dominate and that sharp phase transitions
will be more difficult to produce than in the $p=4$ case.    

\newsec{Conclusions}

\noindent

In this paper we have studied some simple multitrace deformations of the basic
non-supersymmetric QCD model in \refs\rwitthp, as well as its generalizations
to less than four dimensions. 
In particular we have considered deformations by a non-linear function of
the Lagrangian operator.

Our main result is the emergence of  new types of 
``multicritical" behaviour, similar in many ways to those studied in
the context of matrix models \refs\roldmm.   
There appear various competing master fields whose dynamics yields new
examples of large-$N$ phase transitions. It turns out that the dynamical effect
of multitrace deformations is particularly strong in the supergravity approximation
to the AdS/CFT master field. 

These results suggest various avenues for further research. It would be interesting
to study more examples of  
 large-$N$ phase transitions induced by multitrace deformations. 
Eventually, these phase transitions should be related to the breakdown of
string perturbation theory in the geometrical description of the large-$N$ 
master field. Another interesting question is the effect of multitrace
deformations on other large-$N$ phase transitions that have been identified
in single-trace models, in particular, the phase transitions associated
to theta-dependence in \refs\rwitheta\ or those related
to finite-size effects, as in \refs{\rHPage, \rwitthp, \rthresholds}.

\newsec{Acknowledgements}

\noindent

I would like to thank Zack Guralnik and Assaf Shomer  for useful discussions, and
Roberto Emparan for a crucial question.

\baselineskip 12pt

\listrefs

\bye